for coding. For example, for any integer $i \geq 0$ and for any real number $t > 0$, there exists a network such that

$$\mathcal{C}_0^{\text{uniform}} = \mathcal{C}_1^{\text{uniform}} = \cdots = \mathcal{C}_i^{\text{uniform}}$$
$$\mathcal{C}_0^{\text{average}} = \mathcal{C}_1^{\text{average}} = \cdots = \mathcal{C}_i^{\text{average}}$$
$$\mathcal{C}_{i+1}^{\text{uniform}} - \mathcal{C}_i^{\text{uniform}} > t$$
$$\mathcal{C}_{i+1}^{\text{average}} - \mathcal{C}_i^{\text{average}} > t.$$

In Theorem III.2, the existence of networks that achieve prescribed rational-valued node-limited capacity functions was established. It is known in general that not all networks necessarily achieve their capacities [5]. It is presently unknown, however, whether a network coding capacity could be irrational.[5] Thus, we are not presently able to extend Theorem III.2 to real-valued functions. Nevertheless, Theorem III.2 does immediately imply the following asymptotic achievability result for real-valued functions.

*Corollary III.5:* Every monotonically nondecreasing, eventually constant function $f : \mathbb{N} \cup \{0\} \to \mathbb{R}^+$ is the limit of the node-limited uniform and average capacity function of some sequence of directed acyclic networks.

---

[5]It would be interesting to understand whether, for example, a node-limited capacity function of a network could take on some rational and some irrational values, and perhaps achieve some values and not achieve other values. We leave this as an open question.

---

# The Sizes of Optimal $q$-Ary Codes of Weight Three and Distance Four: A Complete Solution

Yeow Meng Chee, Son Hoang Dau, Alan C. H. Ling, and San Ling

*Abstract*—This correspondence introduces two new constructive techniques to complete the determination of the sizes of optimal $q$-ary codes of constant weight three and distance four.

## I. INTRODUCTION

The determination of $A_q(n, d, w)$, the size of an optimal $q$-ary code of length $n$, distance $d$, and constant weight $w$ (all terms are defined in the next section), has been the subject of study [1]–[25] due to several important applications requiring nonbinary alphabets, such as coding for bandwidth-efficient channels and design of oligonucleotide sequences for DNA computing. Recently, Chee and Ling [1] introduced an effective technique for constructing optimal constant-weight $q$-ary codes, which allowed the determination of $A_3(n, 4, 3)$ for all $n$. For $q > 3$, the value of $A_q(n, 4, 3)$ has also been determined, except when $n \geq q$, $n \equiv 4$ or $5 \pmod 6$ [1, Th. 13]. Define the equation shown at the bottom of the next page. The upper bound

$$A_q(n, 4, 3) \leq \min\left\{U_q(n), \binom{n}{3}\right\} \quad (1)$$

has been established in [1 Th. 12]. In each case where the value of $A_q(n, 4, 3)$ has been determined, it is found to meet this upper bound [1, Ths. 13 and 14].

In this correspondence, we determine $A_q(n, 4, 3)$ completely, showing that it meets the upper bound (1) in all cases. First, we extend the technique of [1] to work with large sets with holes. This allows the determination of $A_q(n, 4, 3)$ when $n \equiv 4 \bmod 6$ and $q \leq n$, or when $n \equiv 5 \bmod 6$ and $q \leq n - 1$. A novel method based on sequences is then used to determine $A_q(n, 4, 3)$ for the remaining cases when $n = q$.

## II. DEFINITIONS AND NOTATIONS

The set of integers $\{1, \ldots, n\}$ is denoted by $[n]$. For $q$ a positive integer, we denote the ring $\mathbb{Z}/q\mathbb{Z}$ by $\mathbb{Z}_q$. The set of all nonzero elements of $\mathbb{Z}_q$ is denoted $\mathbb{Z}_q^*$. The $i$th coordinate of a vector $\mathbf{u}$ is denoted by $\mathbf{u}_i$,

Manuscript received September 10, 2007; revised November 11, 2007. The research of Y. M. Chee and S. Ling was supported in part by the Singapore Ministry of Education under Research Grant T206B2204. This work of A. C. H. Ling was done while he was on sabbatical leave at the Division of Mathematical Sciences, School of Physical and Mathematical Sciences, Nanyang Technological University, Singapore 637616, Singapore.

Y. M. Chee is with the Interactive Digital Media R&D Program Office, Media Development Authority, Singapore 179369, Singapore. He is also with the Division of Mathematical Sciences, School of Physical and Mathematical Sciences, Nanyang Technological University, Singapore 637616, Singapore, and the Department of Computer Science, School of Computing, National University of Singapore, Singapore 117590, Singapore (e-mail: ymchee@alumni.uwaterloo.ca).

S. H. Dau and S. Ling are with the Division of Mathematical Sciences, School of Physical and Mathematical Sciences, Nanyang Technological University, Singapore 637616, Singapore (e-mail: lingsan@ntu.edu.sg).

A. C. H. Ling is with the Department of Computer Science, University of Vermont, Burlington, VT 05405 USA (e-mail: aling@emba.uvm.edu).

$i \geq 1$. For $u \in \mathbb{Z}^n$ and positive integers $i$ and $j$, $1 \leq i < j \leq n$, the vector $(u_i, u_{i+1}, \ldots, u_j)$ is denoted $u_{[i,j]}$.

For a vector $u \in \mathbb{Z}^n$ and positive integer $k$, $u + k$ denotes the vector $(u_1 + k, u_2 + k, \ldots, u_n + k) \in \mathbb{Z}^n$, and $u \bmod k$ denotes the vector $(u_1 \bmod k, u_2 \bmod k, \ldots, u_n \bmod k) \in (\mathbb{Z}_k)^n$.

The $q$-ary Hamming $n$-space is the set $\mathcal{H}_q(n) = (\mathbb{Z}_q)^n$ endowed with the Hamming distance metric $d_H$ defined as follows:

$$d_H(u, v) = |\{i \in [n] : u_i \neq v_i\}|,$$

the number of coordinates where $u$ and $v$ differ. The *Hamming weight* of a vector $u \in \mathcal{H}_q(n)$ is the quantity $d_H(u, 0)$, the number of nonzero coordinates of $u$. The *support* of $u$ is defined to be the set $\mathrm{supp}(u) = \{i \in [n] : u_i \neq 0\}$. In other words, the Hamming weight of $u$ is the size of the support of $u$. The set of all elements in $\mathcal{H}_q(n)$ having Hamming weight $w$ is denoted $\mathcal{H}_q(n, w)$. A $q$-ary code of length $n$, distance $d$ and (constant) weight $w$, denoted $(n, d, w)_q$-*code*, is a nonempty set $\mathcal{C} \subseteq \mathcal{H}_q(n, w)$ such that $d_H(u, v) \geq d$ for all $u, v \in \mathcal{C}$, $u \neq v$. The elements of $\mathcal{C}$ are called *codewords*.

The number of codewords in an $(n, d, w)_q$-code is called the *size* of the code. The maximum size of an $(n, d, w)_q$-code is denoted $A_q(n, d, w)$. An $(n, d, w)_q$-code having $A_q(n, d, w)$ codewords is said to be *optimal*.

Given a finite set $X$ and a nonnegative integer $k$, the set of all $k$-subsets of $X$ is denoted $\binom{X}{k}$. A *set system* is a pair $(X, \mathcal{A})$, where $X$ is a finite set of *points* and $\mathcal{A} \subseteq 2^X$, whose elements are called *blocks*. The *order* of the set system is $|X|$, the number of points. For a set of nonnegative integers $K$, a set system $(X, \mathcal{A})$ is said to be $K$-*uniform* if $|A| \in K$ for all $A \in \mathcal{A}$.

A $t$-*wise balanced design*, denoted $t$BD, is a set system $(X, \mathcal{A})$ with the property that every $T \in \binom{X}{t}$ is contained in exactly one block of $\mathcal{A}$. If the $t$BD is $K$-uniform and of order $n$, then we also denote it by $t\mathrm{BD}(n, K)$. A $t\mathrm{BD}(n, \{k\})$ is also commonly known as a *Steiner system*. In particular, a $2\mathrm{BD}(n, \{3\})$ is a Steiner triple system of order $n$.

### III. AN APPLICATION OF LARGE SETS WITH HOLES

Chee and Ling [1] used large sets of Steiner triple systems to determine $A_q(n, 4, 3)$ for $n \equiv 0, 1, 2,$ or $3 \bmod 6$. In this section, we utilize large sets with holes, a useful concept introduced by Teirlinck [26], to determine $A_q(n, 4, 3)$ for $n \equiv 5 \bmod 6$.

*Definition 1:* A large set $\mathrm{LS}(t, (k, K), n)$ is a set $\{(X, \mathcal{A}_r) : r \in R\}$ of $t\mathrm{BD}(n, K)$ such that
1) $(X, \cup_{r \in R} \mathcal{A}_r)$ is a $k\mathrm{BD}(n, K)$; and
2) for each $A \in \cup_{r \in R} \mathcal{A}_r$, there are exactly $\binom{|A|-t}{k-t}$ elements $r \in R$ such that $A \in \mathcal{A}_r$.

Note that in Definition 1, $\cup_{r \in R} \mathcal{A}_r$ denotes the ordinary set union, and not multiset union.

It is known that an $\mathrm{LS}(t, (k, K), n)$ contains $\binom{n-t}{k-t}$ $t\mathrm{BD}(n, K)$ [26, Prop. 1.1]. Teirlinck [26] established a number of existence results for $\mathrm{LS}(t, (k, K), n)$. In particular, the following was obtained.

*Theorem 1 (Teirlinck [26, Prop. 3.2]):* An $\mathrm{LS}(2, (3, \{3, 5\}), n)$ exists if and only if $n \geq 3$ is odd and $n \neq 7$.

When $n \equiv 5 \bmod 6$, $n \geq 5$, the $\mathrm{LS}(2, (3, \{3, 5\}), n)$ that Teirlinck constructed [26, Construction 3.1] in the proof of Theorem 1 has the property that each $2\mathrm{BD}(n, \{3, 5\})$ in the large set contains exactly one block of size five. Consider such an $\mathrm{LS}(2, (3, \{3, 5\}), n)$, say $\mathcal{L} = \{([n], \mathcal{A}_r) : r \in [n-2]\}$. Each $([n], \mathcal{A}_r)$, $r \in [n-2]$, is a $2\mathrm{BD}(n, \{3, 5\})$ containing exactly one block of size five and hence $\frac{1}{3}\left(\binom{n}{2} - 10\right)$ blocks of size three. By the definition of $\mathrm{LS}(2, (3, \{3, 5\}), n)$, each block of size three in $\cup_{r \in [n-2]} \mathcal{A}_r$ appears in exactly one $2\mathrm{BD}(n, \{3, 5\})$ of the large set and each block of size five in $\cup_{r \in [n-2]} \mathcal{A}_r$ appears in exactly three $2\mathrm{BD}(n, \{3, 5\})$ of the large set. Note also that any two blocks in $\cup_{r \in [n-2]} \mathcal{A}_r$ intersect in at most two points, since $([n], \cup_{r \in [n-2]} \mathcal{A}_r)$ is a 3BD.

Let $\mathcal{F} = \{F_1, \ldots, F_{(n-2)/3}\}$ be the set of all blocks of size five in $\cup_{r \in [n-2]} \mathcal{A}_r$. Define for each $i \in [(n-2)/3]$

$$\mathcal{P}_i = \{([n], \mathcal{A}_r) : F_i \in \mathcal{A}_r, R \in [n-2]\}.$$

Then it is easy to see that $\mathcal{P}_i$, $i \in [(n-2)/3]$, are mutually disjoint, and each $\mathcal{P}_i$ contains precisely three $2\mathrm{BD}(n, \{3, 5\})$. Hence, $\mathcal{F}$ induces a partition of $\mathcal{L}$ as follows:

$$\mathcal{L} = \cup_{i=1}^{(n-2)/3} \mathcal{P}_i.$$

We assume without loss of generality that $([n], \mathcal{A}_{3i-2})$, $([n], \mathcal{A}_{3i-1})$, $([n], \mathcal{A}_{3i}) \in \mathcal{P}_i$, for $i \in [(n-2)/3]$.

Let $2 \leq q \leq n - 1$, $\alpha = \lfloor (q-1)/3 \rfloor$, and $\beta = q - 1 - 3\alpha$, so that $q - 1 = 3\alpha + \beta$. For each $r \in [q-1]$, let $\mathcal{C}_r$ be the set of all codewords $u \in \{0, r\}^n$ of weight three such that $\mathrm{supp}(u) \in \mathcal{A}_r$. Further, for each $F_i$, $i \in [\alpha]$, let $\mathcal{C}'_i$ be an optimal $(5, 4, 3)_4$-code on the alphabet set $\{0, 3i-2, 3i-1, 3i\}$ so that $\mathrm{supp}(u) \subset F_i$ for each $u \in \mathcal{C}'_i$. Finally, if $\beta \geq 1$, let $\mathcal{C}'_{\alpha+1}$ be an optimal $(5, 4, 3)_{\beta+1}$-code on the alphabet set $\{3\alpha+1, \ldots, 3\alpha+\beta\} \cup \{0\}$ so that $\mathrm{supp}(u) \subset F_{\alpha+1}$ for each $u \in \mathcal{C}'_{\alpha+1}$. For convenience, define $\mathcal{C}'_{\alpha+1} = \varnothing$ if $\beta = 0$.

It is obvious from its construction that

$$\mathcal{C} = \left(\bigcup_{i=1}^{q-1} \mathcal{C}_i\right) \cup \left(\bigcup_{i=1}^{\alpha+1} \mathcal{C}'_i\right)$$

is a $q$-ary code of length $n$ and weight three. We claim that $\mathcal{C}$ is in fact an optimal $(n, 4, 3)_q$-code. Indeed, suppose $u, v \in \mathcal{C}$ are distinct.

- If $u, v \in \cup_{i=1}^{q-1} \mathcal{C}_i$, we have $d_H(u, v) \geq 4$ since if $\mathrm{supp}(u)$ and $\mathrm{supp}(v)$ are two blocks from the same $2\mathrm{BD}(n, \{3, 5\})$, then they intersect in at most one point, and if $\mathrm{supp}(u)$ and $\mathrm{supp}(v)$ are two blocks from different $2\mathrm{BD}(n, \{3, 5\})$, then they intersect in at most two points but $u, v$ must differ in value in those corresponding coordinates.
- If $u, v \in \cup_{i=1}^{\alpha+1} \mathcal{C}'_i$, we have $d_H(u, v) \geq 4$ since if $u, v \in \mathcal{C}'_i$, for some $i$, then $d_H(u, v) \geq 4$ follows from the fact that $\mathcal{C}'_i$ is a code of distance four, and if $u \in \mathcal{C}'_i$, $v \in \mathcal{C}'_j$ for $i \neq j$, then $\mathrm{supp}(u)$ and $\mathrm{supp}(v)$ intersect in at most two points since $|F_i \cap F_j| \leq 2$, but $u, v$ must differ in value in those corresponding coordinates.
- If $u \in \cup_{i=1}^{q-1} \mathcal{C}_i$ and $v \in \cup_{i=1}^{\alpha+1} \mathcal{C}'_i$, we have $d_H(u, v) \geq 4$ since in the case when $u \in \mathcal{C}_{3i-2} \cup \mathcal{C}_{3i-1} \cup \mathcal{C}_{3i}$ and $v \in \mathcal{C}'_i$, we have $|\mathrm{supp}(u) \cap \mathrm{supp}(v)| \leq 1$, and in the case when $u \in$

$$U_q(n) = \begin{cases} \left\lfloor \frac{(q-1)n}{3} \left\lfloor \frac{n-1}{2} \right\rfloor \right\rfloor - 1, & \text{if } n \equiv 5 \pmod{6} \text{ and } q \not\equiv 1 \pmod{3} \\ \left\lfloor \frac{(q-1)n}{3} \left\lfloor \frac{n-1}{2} \right\rfloor \right\rfloor, & \text{otherwise.} \end{cases}$$

$\mathcal{C}_{3i-2} \cup \mathcal{C}_{3i-1} \cup \mathcal{C}_{3i}$ and $\mathsf{v} \in \mathcal{C}'_j$ ($i \neq j$), $\mathrm{supp}(\mathsf{u})$ and $\mathrm{supp}(\mathsf{v})$ intersect in at most two points and $\mathsf{u}, \mathsf{v}$ must differ in value in those corresponding coordinates.

Hence, we conclude that $\mathcal{C}$ is an $(n,4,3)_q$-code. What remains is for us to compute the size of $\mathcal{C}$. We require the sizes of optimal $(5,4,3)_q$-codes, for $q \in \{2,3,4\}$ (which has been shown to take on the value $U_q(5)$ in [1]).

When $q - 1 \equiv 0 \bmod 3$

$$|\mathcal{C}| = \sum_{i=1}^{q-1} |\mathcal{C}_i| + \sum_{i=1}^{\alpha} A_4(5,4,3)$$
$$= (q-1)\frac{1}{3}\left(\binom{n}{2} - 10\right) + 10\left(\frac{q-1}{3}\right)$$
$$= \frac{(q-1)n(n-1)}{6}$$
$$= U_q(n).$$

When $q - 1 \equiv 1 \bmod 3$

$$|\mathcal{C}| = \sum_{i=1}^{q-1} |\mathcal{C}_i| + \sum_{i=1}^{\alpha} A_4(5,4,3) + A_2(5,4,3)$$
$$= (q-1)\frac{1}{3}\left(\binom{n}{2} - 10\right) + 10\left(\frac{q-2}{3}\right) + 2$$
$$= \frac{(q-1)n(n-1)}{6} - \frac{4}{3}$$
$$= U_q(n).$$

When $q - 1 \equiv 2 \bmod 3$

$$|\mathcal{C}| = \sum_{i=1}^{q-1} |\mathcal{C}_i| + \sum_{i=1}^{\alpha} A_4(5,4,3) + A_3(5,4,3)$$
$$= (q-1)\frac{1}{3}\left(\binom{n}{2} - 10\right) + 10\left(\frac{q-3}{3}\right) + 5$$
$$= \frac{(q-1)n(n-1)}{6} - \frac{5}{3}$$
$$= U_q(n).$$

Therefore, $\mathcal{C}$ is an optimal $(n,4,3)_q$-code.

We can now state the following.

*Theorem 2:* $A_q(n,4,3) = U_q(n)$ for $n \equiv 5 \bmod 6$ and $2 \leq q \leq n - 1$.

*Corollary 1:* $A_q(n,4,3) = U_q(n)$ for $n \equiv 4 \bmod 6$ and $2 \leq q \leq n$.

*Proof:* If $n \equiv 4 \bmod 6$ and $2 \leq q \leq n$, consider an optimal $(n+1,4,3)_q$-code $\mathcal{C}$ of size $U_q(n+1)$. The total number of nonzero coordinates among all the $U_q(n+1)$ codewords is $3U_q(n+1)$, since the weight of each codeword is three. Hence there must exist $i$ such that

$$|\{\mathsf{u} \in \mathcal{C} : \mathsf{u}_i \neq 0\}| \leq \left\lfloor \frac{3U_q(n+1)}{n+1} \right\rfloor$$
$$= \begin{cases} \frac{(q-1)n}{2} - 1, & \text{if } q \equiv 0 \text{ or } 2 \bmod 3 \\ \frac{(q-1)n}{2}, & \text{if } q \equiv 1 \bmod 3. \end{cases}$$

Shorten the code $\mathcal{C}$ at coordinate $i$ to obtain an $(n,4,3)_q$-code. This will remove at most $(q-1)n/2$ or $(q-1)n/2 - 1$ codewords from $\mathcal{C}$, depending on whether $q \equiv 1 \bmod 3$ or otherwise, so that the $(n,4,3)_q$-code we obtain has size at least

$$\begin{cases} U_q(n+1) - \frac{(q-1)n}{2}, & \text{if } q \equiv 1 \bmod 3 \\ U_q(n+1) - \left(\frac{(q-1)n}{2} - 1\right), & \text{if } q \equiv 0 \text{ or } 2 \bmod 3. \end{cases}$$

In each case, this size evaluates to $U_q(n)$, proving that the $(n,4,3)_q$-code thus obtained is optimal. $\square$

At this point, the only values of $A_q(n,4,3)$ that are unknown are for $q = n \equiv 5 \bmod 6$. In Section IV, we settle this problem more generally by constructing optimal $(q,4,3)_q$-codes for all $q \geq 3$ using a construction based on sequences.

## IV. THE VALUE OF $A_q(q,4,3)$

It is known [1] that

$$A_q(q,4,3) \leq \binom{q}{3}. \tag{2}$$

Partial progress on the determination of $A_q(q,4,3)$ was obtained in [1]. This can be summarized as follows.

*Theorem 3 (Chee and Ling [1, Ths. 13 and 14]):*
1) $A_q(q,4,3) = \binom{q}{3}$ when $q \equiv 0,1,2,$ or $3 \pmod{6}$;
2) $A_q(q,4,3) = \binom{q}{3}$ when $q$ is the power of an odd prime.

The proof of Theorem 3 given in [1] relied on an unpublished result of Ding *et al.* [2]. In this section, we establish a more general result on $A_q(q,4,3)$ that is self-contained. In particular, we prove the following.

*Theorem 4:* $A_q(q,4,3) = \binom{q}{3}$ for all $q \geq 3$.

### A. The Construction Method

The elements of $\binom{[n]}{k}$ can be ordered using the *lexicographic order* $\prec$ defined below.

*Definition 2:* For distinct $A, B \in \binom{[n]}{k}$, $A \prec B$ if and only if $\min\{i : i \in A \Delta B\} \in A$.

For $A \in \binom{[n]}{k}$, let $\mathrm{rank}(A)$ denote the position of $A$ in the lexicographic ordering of $\binom{[n]}{k}$; hence, $\mathrm{rank}(\cdot)$ is a bijection

$$\mathrm{rank} : \binom{[n]}{k} \to \left[\binom{n}{k}\right].$$

It is well known (see, for example, [27]) that for $1 \leq t_1 < t_2 < \cdots < t_k \leq n$, we have

$$\mathrm{rank}(\{t_1, t_2, \ldots, t_k\}) = 1 + \sum_{i=1}^{k} \sum_{j=t_{i-1}+1}^{t_i - 1} \binom{n-j}{k-i} \tag{3}$$

where $t_0 = 0$.

Let $\mathsf{M}(n)$ denote the $\binom{n}{3} \times n$ $\{0,1\}$-matrix whose rows are the elements of $\mathcal{H}_2(n,3)$, whose supports are in (ascending) lexicographic order. Let $\mathsf{s} \in (\mathbb{Z}_q^*)^{\binom{n-1}{2}}$ be a $q$-ary sequence of length $\binom{n-1}{2}$ comprising symbols from $\mathbb{Z}_q^*$. We *fill* each column of $\mathsf{M}(n)$ with $\mathsf{s}$ as follows. We traverse the entries of each column in a top-down manner and replace the nonzero elements of the column by the elements of $\mathsf{s}$ in order. More precisely, when filling the $j$th column of $\mathsf{M}(n)$ with $\mathsf{s}$, let $i_1 < i_2 < \cdots < i_{\binom{n-1}{2}}$ be the row indices so that $\mathsf{M}(n)_{i_t,j}$ is nonzero, $t \in \left[\binom{n-1}{2}\right]$. We then replace the entry in $\mathsf{M}(n)_{i_t,j}$ by $\mathsf{s}_t$, $t \in \left[\binom{n-1}{2}\right]$. The resulting matrix is denoted by $\mathsf{M}(n,\mathsf{s})$. It is obvious that the set of rows of $\mathsf{M}(n,\mathsf{s})$ forms a $q$-ary code of constant weight three having size $\binom{n}{3}$. We call this code the code of $\mathsf{M}(n,\mathsf{s})$. The distance of this code would depend on the sequence $\mathsf{s}$. We show in the next section that it is possible to design a $q$-ary sequence $\mathsf{y}(q)$ so that the code of $\mathsf{M}(q,\mathsf{y}(q))$ has distance four.

*Example 1:* Let $\mathsf{s} = (1, 2, 3, 3, 4, 1) \in (\mathbb{Z}_5^*)^6$. Then we have

$$\mathsf{M}(5) = \begin{bmatrix} 1 & 1 & 1 & 0 & 0 \\ 1 & 1 & 0 & 1 & 0 \\ 1 & 1 & 0 & 0 & 1 \\ 1 & 0 & 1 & 1 & 0 \\ 1 & 0 & 1 & 0 & 1 \\ 1 & 0 & 0 & 1 & 1 \\ 0 & 1 & 1 & 1 & 0 \\ 0 & 1 & 1 & 0 & 1 \\ 0 & 1 & 0 & 1 & 1 \\ 0 & 0 & 1 & 1 & 1 \end{bmatrix},$$

$$\mathsf{M}(5,\mathsf{s}) = \begin{bmatrix} 1 & 1 & 1 & 0 & 0 \\ 2 & 2 & 0 & 1 & 0 \\ 3 & 3 & 0 & 0 & 1 \\ 3 & 0 & 2 & 2 & 0 \\ 4 & 0 & 3 & 0 & 2 \\ 1 & 0 & 0 & 3 & 3 \\ 0 & 3 & 3 & 3 & 0 \\ 0 & 4 & 4 & 0 & 3 \\ 0 & 1 & 0 & 4 & 4 \\ 0 & 0 & 1 & 1 & 1 \end{bmatrix}.$$

The code of $\mathsf{M}(5, \mathsf{s})$ is a $(5, 4, 3)_5$-code of size $\binom{5}{3} = 10$.

*B. Sequence Design*

We call a sequence $\mathsf{s} \in (\mathbb{Z}_q^*)^{\binom{q-1}{2}}$ such that the code of $\mathsf{M}(q, \mathsf{s})$ has distance four a *special sequence*, and denote it by $S(q)$.

If $\mathsf{u}$ and $\mathsf{v}$ are two distinct rows of $\mathsf{M}(q,\mathsf{s})$, then $|\mathrm{supp}(\mathsf{u}) \cap \mathrm{supp}(\mathsf{v})| \in \{0, 1, 2\}$. Futhermore, if $|\mathrm{supp}(\mathsf{u}) \cap \mathrm{supp}(\mathsf{v})| \in \{0, 1\}$, then $d_H(\mathsf{u}, \mathsf{v}) \geq 4$. Hence, $\mathsf{s}$ is a special sequence if and only if $d_H(\mathsf{u}, \mathsf{v}) = 4$ for any two distinct rows $\mathsf{u}$ and $\mathsf{v}$ of $\mathsf{M}(q, s)$ satisfying $|\mathrm{supp}(\mathsf{u}) \cap \mathrm{supp}(\mathsf{v})| = 2$.

For $q \geq 3$, define the sequence

$$\mathsf{x}(q) = \mathsf{x}(q)^{(q-2)}\mathsf{x}(q)^{(q-3)} \cdots \mathsf{x}(q)^{(1)}$$

where

$$\mathsf{x}(q)^{(t)} = \begin{cases} (0, 1, 2, \ldots, q-3), & \text{if } t = q-2 \\ (\mathsf{x}(q)^{(t+1)} + 2)_{[1,t]} \bmod q - 1, & \text{if } t \in [q-3]. \end{cases}$$

Explicitly, we have, for $1 \leq i \leq t \leq q-2$

$$\mathsf{x}(q)_i^{(t)} = 2(q-2-t) + (i-1) \bmod q - 1. \quad (4)$$

Further, define

$$\mathsf{y}(q) = \mathsf{x}(q) + 1.$$

Then $\mathsf{y}(q) \in (\mathbb{Z}_q^*)^{\binom{q-1}{2}}$.

*Example 2:* The following table lists the sequences $\mathsf{y}(q)$, for $3 \leq q \leq 10$:

| $q$ | $\mathsf{y}(q)$ |
|---|---|
| 3 | 1 |
| 4 | 123 |
| 5 | 123341 |
| 6 | 1234345512 |
| 7 | 123453456561123 |
| 8 | 123456345675671712234 |
| 9 | 1234567345678567817812123345 |
| 10 | 123456783456789567891789129123234456 |

We show that $\mathsf{y}(q)$ is a special sequence for all $q \geq 3$.

*Lemma 1:* Let $q \geq 3$ and $\mathsf{A}$ be a $\binom{q}{3} \times q$ matrix such that the supports of its rows are all the elements of $\binom{[q]}{3}$ in lexicographic order. Further, let $2 \leq x \leq q$ and $\mathsf{u}, \mathsf{v}$ be two distinct rows of $\mathsf{A}$ such that $\mathrm{supp}(\mathsf{u}) \cap \mathrm{supp}(\mathsf{v}) = \{1, x\}$. If the first column of $\mathsf{A}$ is filled with $\mathsf{y}(q)$, then $\mathsf{u}_1 \neq \mathsf{v}_1$.

*Proof:* Suppose $\mathrm{supp}(\mathsf{u}) = \{1, x, a\}$ and $\mathrm{supp}(\mathsf{v}) = \{1, x, b\}$, $a, b \neq \in \{1, x\}$. Without loss of generality, assume $a < b$. There are three cases to consider.

When $1 < a < b < x$, we have by (3)

$$\mathrm{rank}(\{1, a, x\}) = \sum_{j=2}^{a-1}(q-j) + (x-a)$$

$$\mathrm{rank}(\{1, b, x\}) = \sum_{j=2}^{b-1}(q-j) + (x-b).$$

If the first column of $\mathsf{A}$ is filled with $\mathsf{y}(q)$, we have $\mathsf{u}_1 = \mathsf{y}(q)_{x-a}^{(q-a)}$ and $\mathsf{v}_1 = \mathsf{y}(q)_{x-b}^{(q-b)}$. Hence, $\mathsf{u}_1 = \mathsf{v}_1$ if and only if $\mathsf{x}(q)_{x-a}^{(q-a)} = \mathsf{x}(q)_{x-b}^{(q-b)}$, which [by (4)] holds if and only if $a = b$. This shows $\mathsf{u}_1 \neq \mathsf{v}_1$.

The cases $1 < a < x < b$ and $1 < x < a < b$ can be dealt with in a similar manner. □

Given an $\binom{n}{3} \times n$ matrix $\mathsf{A}$, such that the supports of its rows are all the elements of $\binom{[n]}{3}$ in lexicographic order, let $\mathsf{A}_j$ denote the matrix obtained by moving column $j$ of $\mathsf{A}$ to the front, where $j \in [n]$. Perform the following *reorder* operation on $\mathsf{A}_j$:

---
Reorder :
Traverse the first column $\mathsf{c}$ of $\mathsf{A}_j$ in a top-down manner. If $\mathsf{c}$ is such that $\mathsf{c}_1, \ldots, \mathsf{c}_{\binom{n-1}{2}} \neq 0$ and $\mathsf{c}_{\binom{n-1}{2}+1}, \ldots, \mathsf{c}_{\binom{n}{3}} = 0$, then stop. Otherwise, let $s = \min\{i : \mathsf{c}_i = 0\}$ and let $t = \min\{i > s : \mathsf{c}_i \neq 0\}$. Move row $t$ of $\mathsf{A}_j$ to the position just before row $s$. Repeat.

---

The resulting matrix is denoted $\mathsf{A}_j'$. We show below that the reorder operation puts the supports of the rows of $\mathsf{A}_j$ into lexicographic order.

*Lemma 2:* If $U, V \in \binom{[n]}{k}$, $U \prec V$, and $x \in U \cap V$, then $U \setminus \{x\} \prec V \setminus \{x\}$.

*Proof:* Since $x \in U \cap V$, $x \neq \in U \Delta V$. Hence, $\min\{i : i \in (U \setminus \{x\})\Delta(V \setminus \{x\})\} = \min\{i : i \in U \Delta V\} \in U$, implying $U \setminus \{x\} \prec V \setminus \{x\}$. □

*Lemma 3:* The supports of the rows of $\mathsf{A}_j'$ are in lexicographic order.

*Proof:* Let $\mathsf{u}$ and $\mathsf{v}$ be rows $i_1$ and $i_2$ of $\mathsf{A}_j'$, $i_1 < i_2$, and let $U = \mathrm{supp}(\mathsf{u}), V = \mathrm{supp}(\mathsf{v})$. We show that $U \prec V$.

If $1 \leq i_1 \leq \binom{n-1}{2}$ and $\binom{n-1}{2}+1 \leq i_2 \leq \binom{n}{3}a$, then by definition of $\mathsf{A}_j'$, we have $1 \in U$ and $1 \neq \in V$. Hence, $\min\{i : i \in U \Delta V\} = 1 \in U$ implying $U \prec V$.

If $1 \leq i_1 < i_2 \leq \binom{n-1}{2}$, then $\mathsf{u}$ and $\mathsf{v}$ corresponds to two rows in $\mathsf{A}$ whose supports contain a common element $j$. By considering the deletion of $j$ from these supports, we see that $U \prec V$ by Lemma 2.

If $\binom{n-1}{2} + 1 \leq i_1 < i_2 \leq \binom{n}{3}$, it is clear that $U \prec V$ since the reorder operation does not change their relative order in $\mathsf{A}$. □

We are now ready to establish:

*Theorem 5:* The sequence $\mathsf{y}(q)$ is a special sequence for all $q \geq 3$.

*Proof:* Let $\mathsf{u}$ and $\mathsf{v}$ be any two distinct rows of $\mathsf{M}(q, \mathsf{y}(q))$ satisfying $|\mathrm{supp}(\mathsf{u}) \cap \mathrm{supp}(\mathsf{v})| = 2$. By a previous comment in Section IV-B, it suffices to show that $d_H(\mathsf{u}, \mathsf{v}) = 4$. Suppose $\mathrm{supp}(\mathsf{u}) \cap \mathrm{supp}(\mathsf{v}) = \{i, j\}$. Then by Lemma 3, $\mathsf{M}_i'(q, \mathsf{y}(q))$ is

a matrix satisfying the hypothesis of Lemma 1. Hence, Lemma 1 implies that $\mathsf{u}_i \neq \mathsf{v}_i$. Similarly, by considering $\mathsf{M}'_j(q, \mathsf{y}(q))$, we have $\mathsf{u}_j \neq \mathsf{v}_j$. This proves $d_H(\mathsf{u}, \mathsf{v}) = 4$. □

This shows that $A_q(q, 4, 3) = \binom{q}{3}$ for all $q \geq 3$. Theorem 4 now follows.

## V. CONCLUSION

In this correspondence, we complete the determination of $A_q(n, 4, 3)$ by employing large sets with holes to construct optimal $(n, 4, 3)_q$-codes for $n \equiv 4$ or $5 \bmod 6$, $n \geq q - 1$, and by using a new technique based on special sequences to construct optimal $(q, 4, 3)_q$-codes. The results of this correspondence combine with those in [1] to give:

Main Theorem: $A_q(n, 4, 3) = \min\{U_q(n), \binom{n}{3}\}$ for all $n$ and $q$.

### ACKNOWLEDGMENT

The authors are grateful to Lingbo Yu for her help in locating literature. The authors are also grateful to the anonymous reviewer for comments that helped improve the presentation of this correspondence.

# Markov Processes Asymptotically Achieve the Capacity of Finite-State Intersymbol Interference Channels

Jiangxin Chen and Paul H. Siegel, *Fellow, IEEE*

*Abstract*—Recent progress in capacity evaluation has made it possible to compute a sequence of lower bounds on the capacity of a finite-state intersymbol-interference (ISI) channel by finding a sequence of optimized Markov input processes with increasing order $r$, for which the state of the process is the previous $r$ input symbols. In this correspondence, we prove that, as the order $r$ goes to infinity, the sequence of optimized Markov sources asymptotically achieves the capacity of the channel. The conclusion is extended to two-dimensional finite-state ISI channels, the binary-symmetric channel (BSC) with constrained inputs, and general indecomposable finite-state channels with a mild constraint.

*Index Terms*—Capacity, finite-state channels, intersymbol interference (ISI) channels, Markov processes, run-length limited constraints, two-dimensional channels.

## I. INTRODUCTION

Magnetic recording channels are generally modeled as finite-state, linear intersymbol-interference (ISI) channels with additive Gaussian noise and a binary input constraint. While the capacity of a general Gaussian linear ISI channel can be evaluated with the water-filling formula [1], a formula for the capacity when the input is constrained to a finite alphabet remains unknown.

Manuscript received December 16, 2004; revised March 3, 2007. This work was supported in part by the National Science Foundation under Grant CCF-0514859 and by the Center for Magnetic Recording Research. The material in this correspondence was presented in part at the IEEE International Symposium on Information Theory, Chicago, IL, June/July 2004.

J. Chen was with the Department of Electrical and Computer Engineering, University of California, San Diego, La Jolla, CA 92093-0407 USA. He is now with the Prediction Company, UBS, Santa Fe, NM, 87505 USA (e-mail: simple_address@yahoo.com).

P. H. Siegel is with the Center for Magnetic Recording Research, University of California, San Diego, La Jolla, CA, 92093-0401 USA (e-mail: psiegel@ucsd.edu).

Communicated by Y. Steinberg, Associate Editor for Shannon Theory.
Digital Object Identifier 10.1109/TIT.2007.915709